# Quantum Machine Learning in Medical Imaging: A Primer for Radiologists

Ruben Ortiz Cordero, MD[1]; Gaven K. Smart, MD[1]; Harshada Gothankar, MSc[2]; Emilio Supsupin, MD[1]; Priya Bhosale, MD[3]; Mayur Virarkar, MD, MBA[1]

[1]Department of Radiology, University of Florida College of Medicine – Jacksonville
[2]Generative AI and Agentic Certifications, Johns Hopkins University
[3]Department of Radiology, MD Anderson Cancer Center

**Corresponding author:** Mayur K. Virarkar, MD, Department of Radiology, UF Health Jacksonville, 655 W. 8th Street, Jacksonville, FL 32209. Email: mayur.virarkar@jax.ufl.edu

## Abstract

Deep learning has transformed radiology, but the computational cost of training and deploying modern models is outpacing what classical hardware can support. Quantum computing has been proposed as one response to this bottleneck: an n-qubit system can represent $2^n$ states simultaneously, although practical speedups require efficient algorithms, data encoding, and measurement strategies rather than superposition alone. This review introduces radiologists to five core quantum concepts, qubits, superposition, entanglement, quantum gates, and quantum circuits, and to the superconducting, trapped-ion, and photonic hardware platforms that define the current noisy intermediate-scale quantum (NISQ) era. We then summarize the quantum machine learning architectures most often applied to medical imaging (quantum support vector machines, quantum neural networks, variational quantum circuits, and quantum convolutional neural networks) and quantum image representations such as the flexible representation of quantum images (FRQI) and the novel enhanced quantum representation (NEQR). Reported applications include brain tumor classification and segmentation on MRI, pneumonia detection on chest radiographs, skin and breast lesion classification, CT and PET reconstruction, and ultrasound denoising. Nearly all published systems are hybrid quantum-classical pipelines evaluated on small, downsampled benchmark datasets or simulators, and clinical deployment is not yet feasible. We conclude with current limitations and translational priorities for radiology.

## Introduction

Over the past decade, artificial intelligence (AI), particularly deep learning, has changed how images are interpreted in radiology. Convolutional neural networks (CNNs) automatically extract hierarchical visual features from image data and have achieved performance comparable to, and sometimes exceeding, that of human experts in classification, segmentation, and detection tasks across CT, MRI, and pathology (1–3). AI-assisted systems have accelerated clinical decision-making in applications that include pneumonia, coronavirus disease 2019 (COVID-19), and stroke (4–6).

Alongside this progress, the computational demands of imaging AI are outpacing traditional hardware. Training models on volumetric MRI and CT examinations requires computing power and memory that grow rapidly with image size, and architectures such as ResNet, DenseNet, and vision transformers carry parameter counts that limit deployment in resource-constrained settings (7–10). Some medical computing problems, radiation therapy plan optimization, molecular simulation, and high-dimensional image analysis, scale so unfavorably that they remain slow or intractable on conventional hardware; quantum annealing has already been applied to intensity-modulated radiation therapy beamlet optimization (11,12).

Quantum computing has emerged as a computational model that can offer exponential speedups for certain classes of problems (8,13). Shor's algorithm factors integers in polynomial time (14), and Grover's algorithm quadratically accelerates unstructured search (15); together, they illustrate that quantum advantage is problem-specific rather than universal. The union of quantum computing and machine learning has produced quantum machine learning (QML) (16,17), which is attracting increasing interest across genomics, drug discovery, diagnostics, and radiotherapy optimization (8,11,18). It has begun to appear in the radiology literature (19). The goal of this review is to provide radiologists with an accessible introduction to the principles of quantum computing and an evidence-focused overview of reported applications in medical imaging analysis.

This article is a narrative primer rather than a systematic review. To identify the literature discussed, we searched PubMed, arXiv, and web-based academic search engines through August 2026, combining quantum computing terms (eg, quantum machine learning, quantum convolutional neural network, quantum annealing, quantum image processing) with medical imaging terms (eg, radiology, MRI, CT, PET, ultrasound), and traced the reference lists of retrieved articles. We prioritized peer-reviewed original studies and reviews applying quantum or quantum-inspired methods to medical images, supplemented by foundational quantum computing sources; all citations were verified against publisher records.

## Fundamentals of Quantum Computing

Quantum computing is a computational approach based on the laws of quantum mechanics rather than binary logic (11,20). This section contrasts classical and quantum computation, defines the key quantum concepts, and summarizes current hardware platforms, without assuming a background in physics or advanced mathematics.

### *Classical Versus Quantum Computing*

The capacity of a classical system scales linearly with its number of bits. In contrast, the state space of a quantum system grows exponentially with each added qubit: an n-qubit register can hold $2^n$ basis states in superposition. Measurement, however, collapses the register to a single outcome, so practical advantage requires interference, efficient circuit design, and problem structure suited to quantum algorithms, not merely the ability to represent $2^n$ possibilities (7–9,21). Figure 1 compares the fundamental computational properties of classical and quantum systems.

### *Core Quantum Computing Concepts*

Five concepts are essential for understanding quantum algorithms in medical imaging: qubits, superposition, entanglement, quantum gates, and quantum circuits (Table 1, Figure 2). A qubit, unlike a classical bit fixed at 0 or 1, can exist as a weighted combination of both states simultaneously (17). This property, called superposition, lets a quantum system hold multiple states in parallel until measurement collapses it to a single classical outcome (11,13). Entanglement is a nonclassical correlation that links the states of two or more qubits, allowing entangled registers to represent complex interdependencies in imaging data compactly (13,20). Quantum gates, unitary, reversible analogs of classical logic gates, are the building blocks from which circuits are constructed (7,17,20). A quantum circuit is a structured sequence of gates, typically comprising a data-encoding layer, a parameterized transformation layer, and a measurement layer (22). Because measurement is probabilistic, circuits are run many times to estimate expectation values, which are then used to produce the final diagnostic prediction (7).

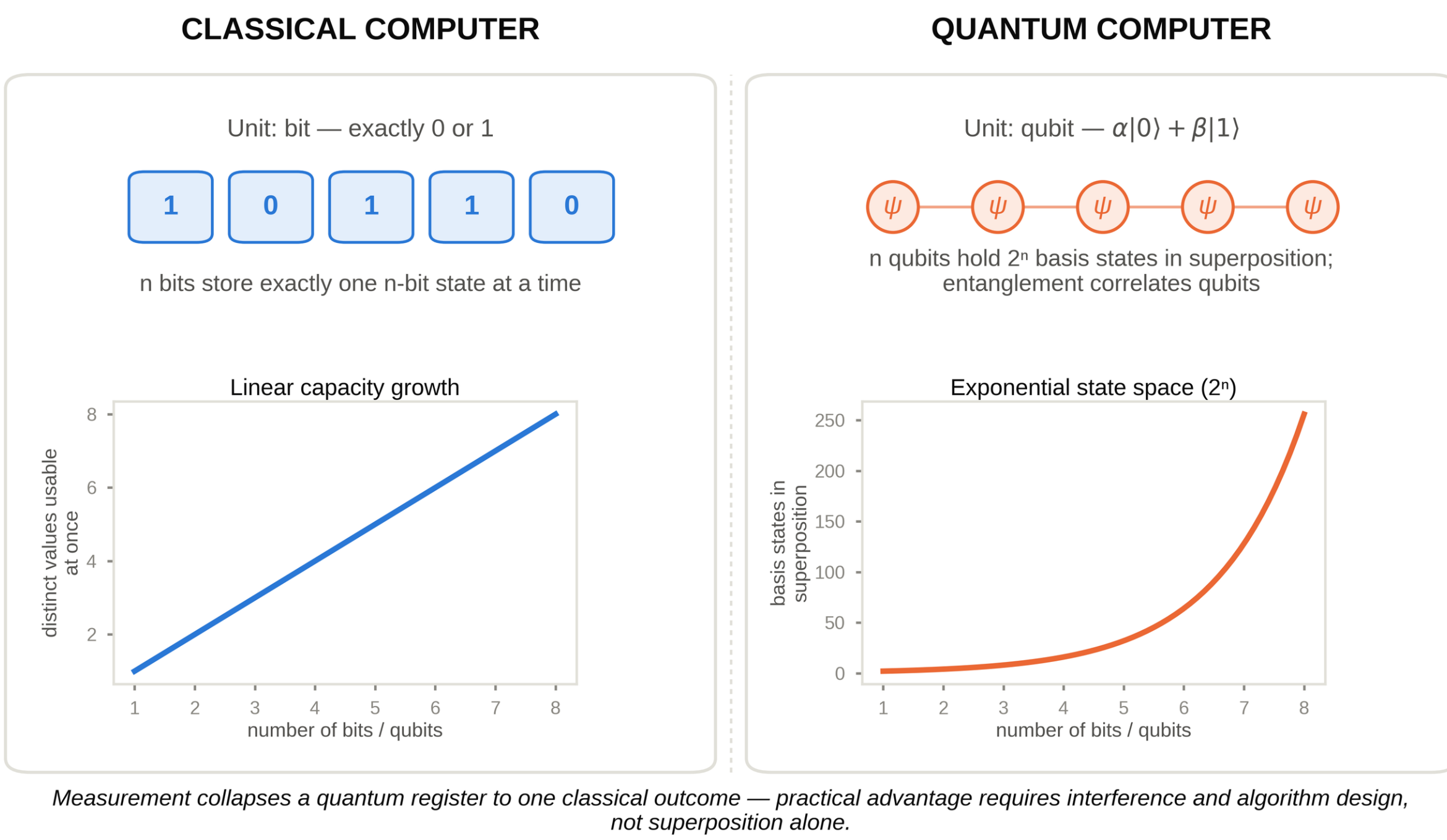

**Figure 1.** Classical versus quantum computation. A classical register of n bits occupies exactly one of its $2^n$ possible states at any moment, and capacity scales linearly with bit count. An n-qubit quantum register can hold all $2^n$ basis states in superposition and represent correlations through entanglement. Still, measurement returns a single classical outcome, so practical advantage depends on interference and algorithm design.

**Table 1.** Essential Quantum Computing Concepts for Radiologists

| Concept | Definition | Computational Role | Medical Imaging Relevance |
|---|---|---|---|
| Qubit | Basic quantum information unit; can hold 0, 1, or both simultaneously | Stores exponentially more state information than bits | Compact representation of image data |
| Superposition | A qubit occupies multiple states at once until measured | Evaluates many possibilities in parallel | Parallel exploration of feature combinations |
| Entanglement | Nonclassical correlation linking qubit states | Models complex dependencies compactly | Captures relationships among tumor features |
| Quantum gates | Reversible operations that manipulate qubit states | Building blocks of all quantum algorithms | Implement image transformations and classifiers |
| Quantum circuits | Gate sequences with encoding, transformation, and measurement layers | Complete quantum computation or model | Diagnostic prediction (eg, tumor detection) |

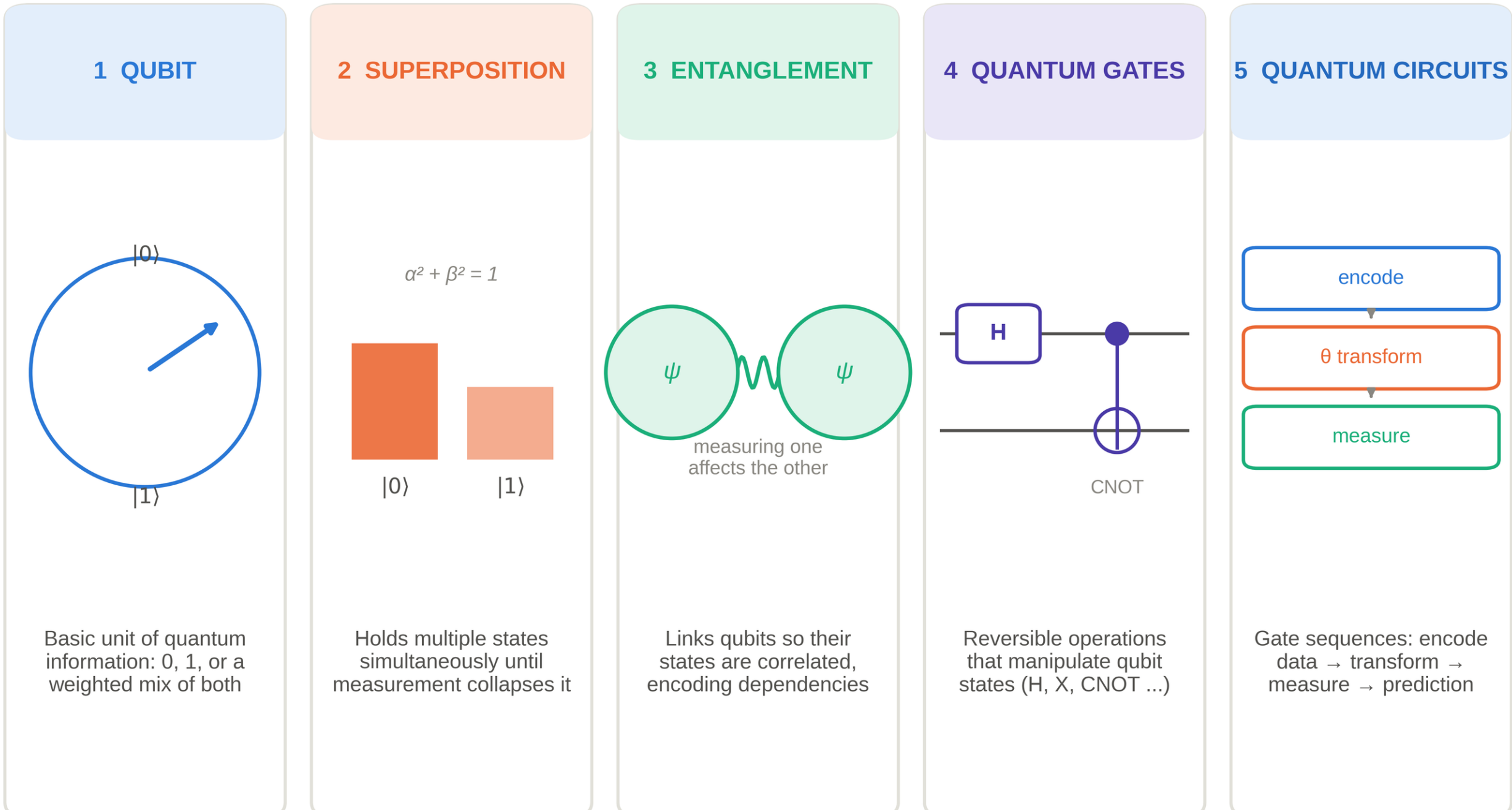

**Figure 2.** Five core quantum computing concepts. A qubit is the basic unit of quantum information. Superposition allows a qubit to occupy weighted combinations of 0 and 1 until measured. Entanglement links qubits so their states are correlated. Quantum gates are reversible operations that manipulate qubit states. Quantum circuits chain gates into data-encoding, parameterized transformation, and measurement layers whose repeated readout yields the prediction.

### *Current Hardware Platforms*

Current devices operate in the noisy intermediate-scale quantum (NISQ) era, with tens to a few hundred qubits and no full error correction (23). Three physical architectures dominate development (Figure 3). Superconducting processors, built from microfabricated circuits cooled to near absolute zero, are the dominant platform; IBM provides cloud access to this hardware, and recent medical image classification benchmarking used IBM's 127-qubit Cleveland system (11,24). Trapped-ion processors suspend charged atoms in electromagnetic fields, offering longer coherence times and distinct noise characteristics; they have been explored in digital health QML studies, although certain parallel operations demonstrated at small scales are not expected to generalize to larger qubit counts (7). Photonic platforms encode information in light and operate at room temperature, but remain difficult to control and are underrepresented in medical imaging research (13).

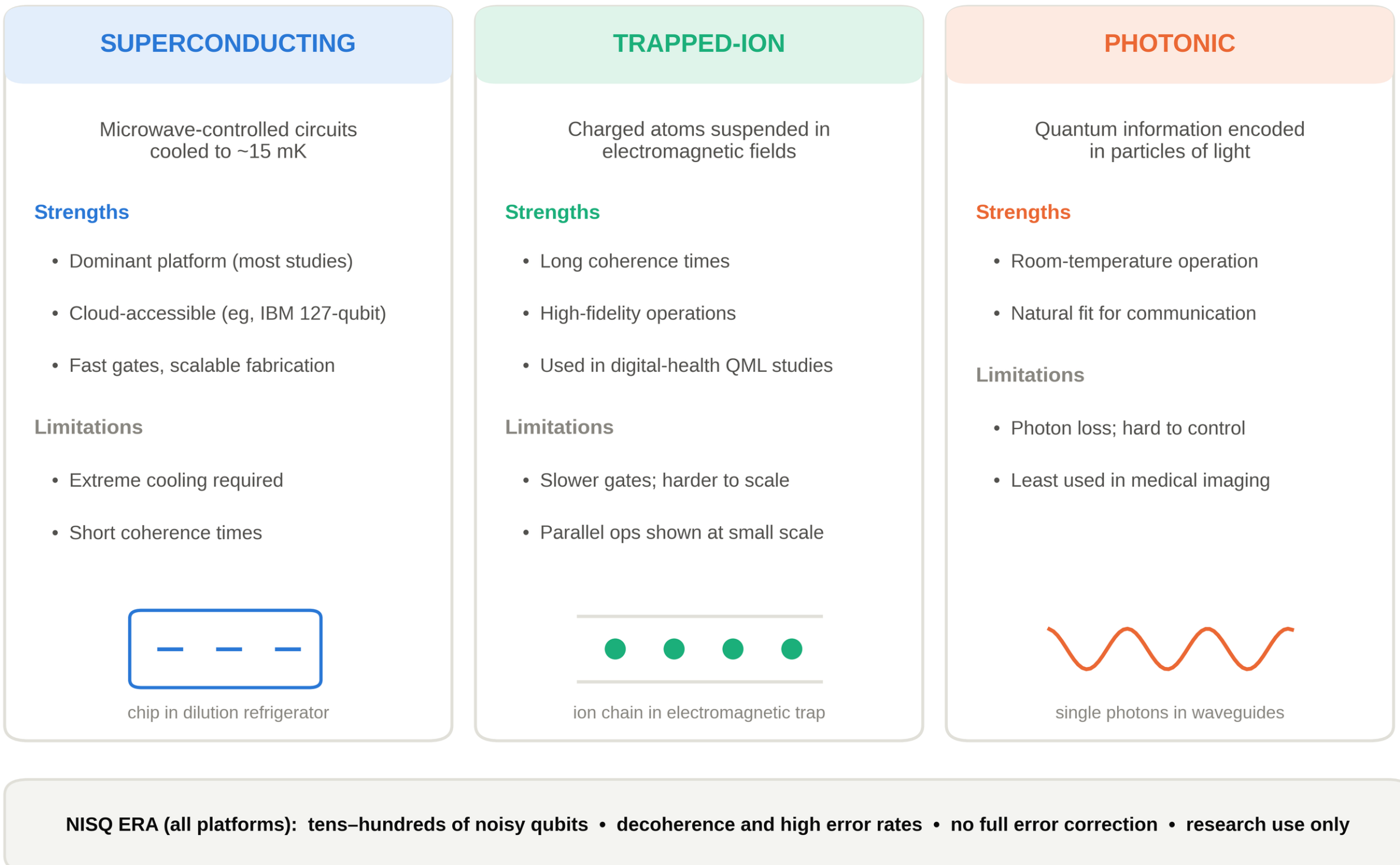

**Figure 3.** Quantum hardware platforms in the noisy intermediate-scale quantum (NISQ) era. Superconducting processors dominate medical imaging research and are cloud accessible. Trapped-ion systems offer longer coherence times at smaller scales. Photonic platforms operate at room temperature but remain difficult to control. All current devices are noisy, have limited qubit counts, and lack full error correction.

All platforms share decoherence, the loss of quantum states through environmental disturbance, which degrades information before a computation completes (11,23). The consequences are measurable: in a benchmark of quantum models for medical image classification on IBM hardware, device noise alone reduced accuracy by up to 15.70 percentage points relative to a noise-free simulation of the same model (24).

## Quantum Machine Learning

QML applies quantum computing to machine learning tasks to improve performance on computationally demanding problems (16,17). In medical imaging research, QML has been used for classification, segmentation, and feature extraction (20,25). The most frequently applied architectures, quantum support vector machines (QSVMs), quantum neural networks, variational quantum circuits (VQCs), and quantum convolutional neural networks (QCNNs), are summarized in Table 2 and Figure 4A (9).

**Table 2.** Quantum Machine Learning Architectures Applied to Medical Imaging

| Architecture | Mechanism | Advantage | Current Limitation |
|---|---|---|---|
| QSVM | Classifies data in a quantum feature space | May separate class boundaries that are hard to model classically | Requires classical preprocessing to reduce input size |
| QNN | Neural network layers built from quantum circuit operations | Explores multiple solutions simultaneously during training | Hybrid designs only; circuit depth limited by NISQ hardware |
| Quanvolutional network | Small quantum circuits act as filters over image patches | Quantum feature maps differ from classical convolution | Operates on small patches |
| VQC | Parameterized circuit trained by a classical optimizer | Most NISQ-compatible; widely used in imaging studies | Trainability limited by noise and barren plateaus |
| QCNN | Convolution-and-pooling structure in the quantum domain | O(log N) trainable parameters | Hybrid designs; classical output layer required |

Note.—NISQ = noisy intermediate-scale quantum, QCNN = quantum convolutional neural network, QNN = quantum neural network, QSVM = quantum support vector machine, VQC = variational quantum circuit.

QSVMs are quantum counterparts to classical support vector machines that learn optimal boundaries between categories, such as malignant and benign (26). They operate in a quantum feature space, an exponentially large representation in which some class boundaries that are difficult to model classically may become separable, and have been used for binary tasks such as pneumonia detection and retinopathy screening (9,26). Quantum neural networks mirror the layered structure of classical networks but replace matrix multiplications with quantum circuit operations (9). VQCs, the most widely used architecture in current research, combine a data-encoding layer, adjustable quantum operations, and a measurement step, with parameters updated by classical optimization during training (27). QCNNs translate the convolutional design that dominates medical image analysis into the quantum domain and can process an N-qubit input with only O(log N) trainable parameters (9,22,28). A related design applies small quantum circuits as feature-extracting filters over image patches within an otherwise classical network (29).

Because NISQ hardware cannot yet support end-to-end quantum pipelines for clinical imaging, most published systems are hybrid: classical networks perform preprocessing and feature extraction, quantum circuits perform feature transformation or classification, and classical optimizers update the quantum parameters during training (Figure 4B) (7,9).

## Quantum Image Processing

Quantum image processing converts conventional image-processing operations to quantum representations of image data: classical images are encoded as quantum states, and quantum algorithms then manipulate, analyze, or protect that information (30,31). In the medical imaging literature, its main uses are image representation, image operations such as edge detection and segmentation, and security methods including encryption and watermarking (Table 3) (31,32).

### *Quantum Image Representation*

Encoding is the first step. The flexible representation of quantum images (FRQI) stores pixel intensity in probability amplitudes and position in qubits, prepared through a constructive polynomial procedure; because amplitudes carry intensity, accurate recovery is probabilistic and requires many repeated measurements (33). The novel enhanced quantum representation (NEQR) instead stores grayscale values in orthogonal basis states of an entangled qubit sequence, permitting direct readout; the original report described lower preparation complexity and higher compression ratios than FRQI (32,34). Figure 5 compares the two encodings. After encoding, operations such as edge detection, segmentation, compression, and classification act on the quantum state rather than on classical pixel arrays, paralleling routine preprocessing and analysis tasks (32–34).

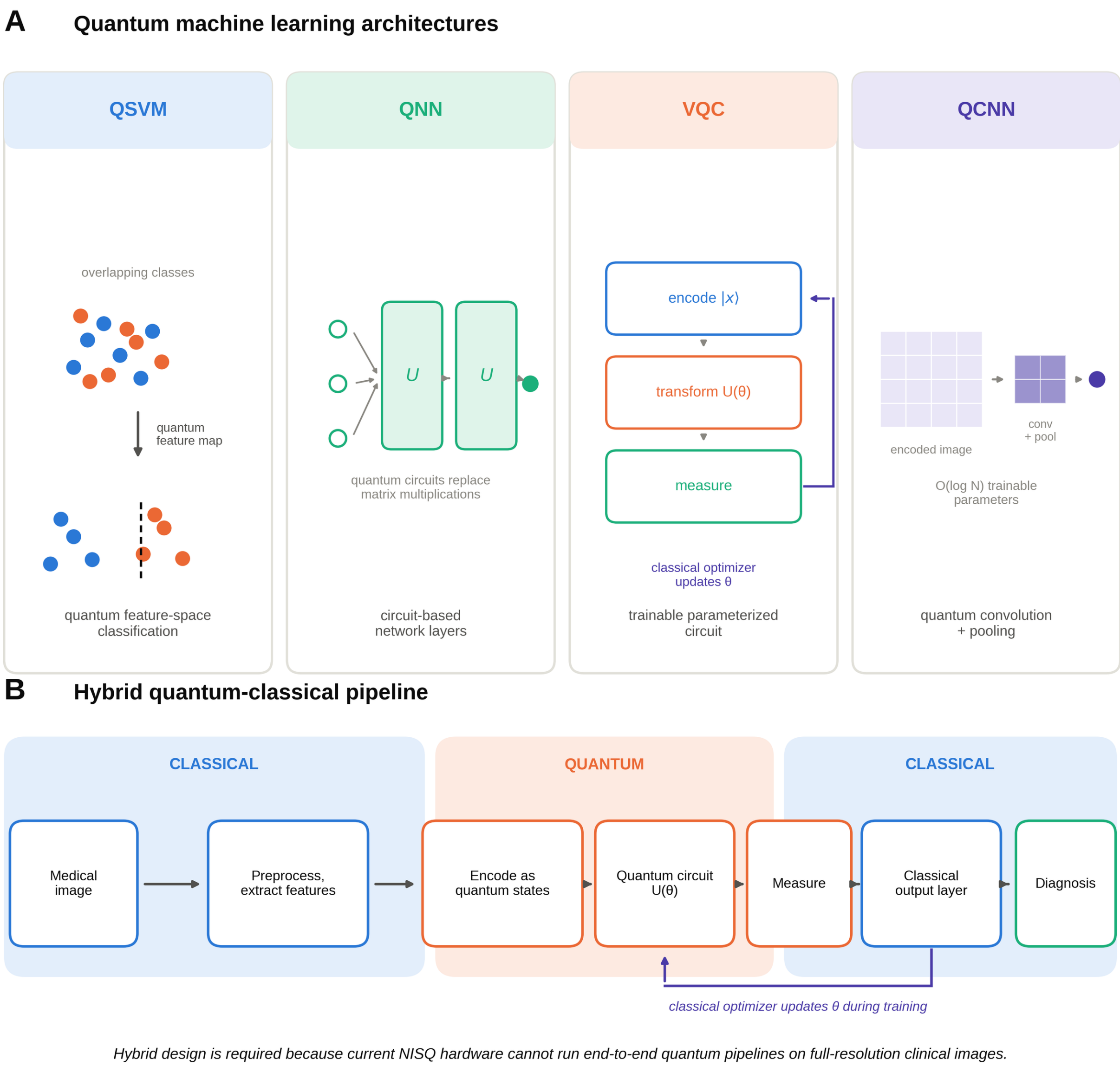

**Figure 4.** Quantum machine learning in practice. (A) The four architectures most used in medical imaging: quantum support vector machines (QSVMs) classify in a quantum feature space; quantum neural networks (QNNs) build layers from circuit operations; variational quantum circuits (VQCs) train encode-transform-measure loops through a classical optimizer; quantum convolutional neural networks (QCNNs) apply convolution and pooling in the quantum domain with O(log N) parameters. (B) The hybrid quantum-classical pipeline used by most published systems comprises classical preprocessing and feature extraction; quantum encoding and transformation; measurement; and a classical output layer, with a classical optimizer updating circuit parameters during training.

**Table 3.** Main Uses of Quantum Image Processing in Medical Imaging

| Category | Purpose | Representative Methods | Relevance |
|---|---|---|---|
| Image representation | Encode classical pixel data as quantum states | FRQI (amplitude encoding); NEQR (basis-state encoding) (33,34) | Foundation for all downstream quantum operations |
| Image operations | Transform the encoded image state | Edge detection, segmentation, compression, classification (32) | Parallels routine preprocessing and analysis tasks |
| Encryption | Prevent unauthorized viewing | Quantum walk and quantum chaotic map key generation (20) | Protects images transmitted over public networks |
| Watermarking | Verify authenticity and ownership | Quantum particle swarm-optimized adaptive embedding (20) | Detects tampering without degrading interpretation |
| Steganography | Conceal existence of sensitive data | Quantum walk embedding (20) | High-capacity covert protection of patient data |

Note.—FRQI = flexible representation of quantum images, NEQR = novel enhanced quantum representation. Numerals refer to the reference list.

### *Quantum Image Security*

With growing reliance on teleradiology, medical images transmitted over public networks are vulnerable to tampering, theft, and unauthorized access (20). Quantum-enabled security offers three complementary mechanisms. Quantum image encryption converts images into a noise-like, visually meaningless format using quantum-generated pseudorandom keys; quantum walks and quantum chaotic maps produce highly unpredictable key sequences that resist histogram and differential attacks, and the quantum no-cloning theorem makes intercepted keys difficult to replicate (20). Quantum watermarking embeds an invisible signature, such as a hospital logo or patient identifier, to verify authenticity and ownership, balancing imperceptibility against robustness; quantum particle swarm optimization has been used to tune embedding parameters adaptively (20). Quantum steganography conceals the very existence of sensitive data within a host image, with quantum walk-based embedding achieving high peak signal-to-noise ratios in experimental validation (20). Together, these methods protect image integrity across storage and transmission.

## Applications in Medical Imaging

Applications of quantum computing in medical imaging remain experimental. Current research supports the exploratory use of AI in classification, segmentation, radiomics, and reconstruction. Still, most studies are hybrid proof-of-concept systems constrained by NISQ hardware and are not ready for clinical deployment (20). Table 4 summarizes reported applications by imaging modality.

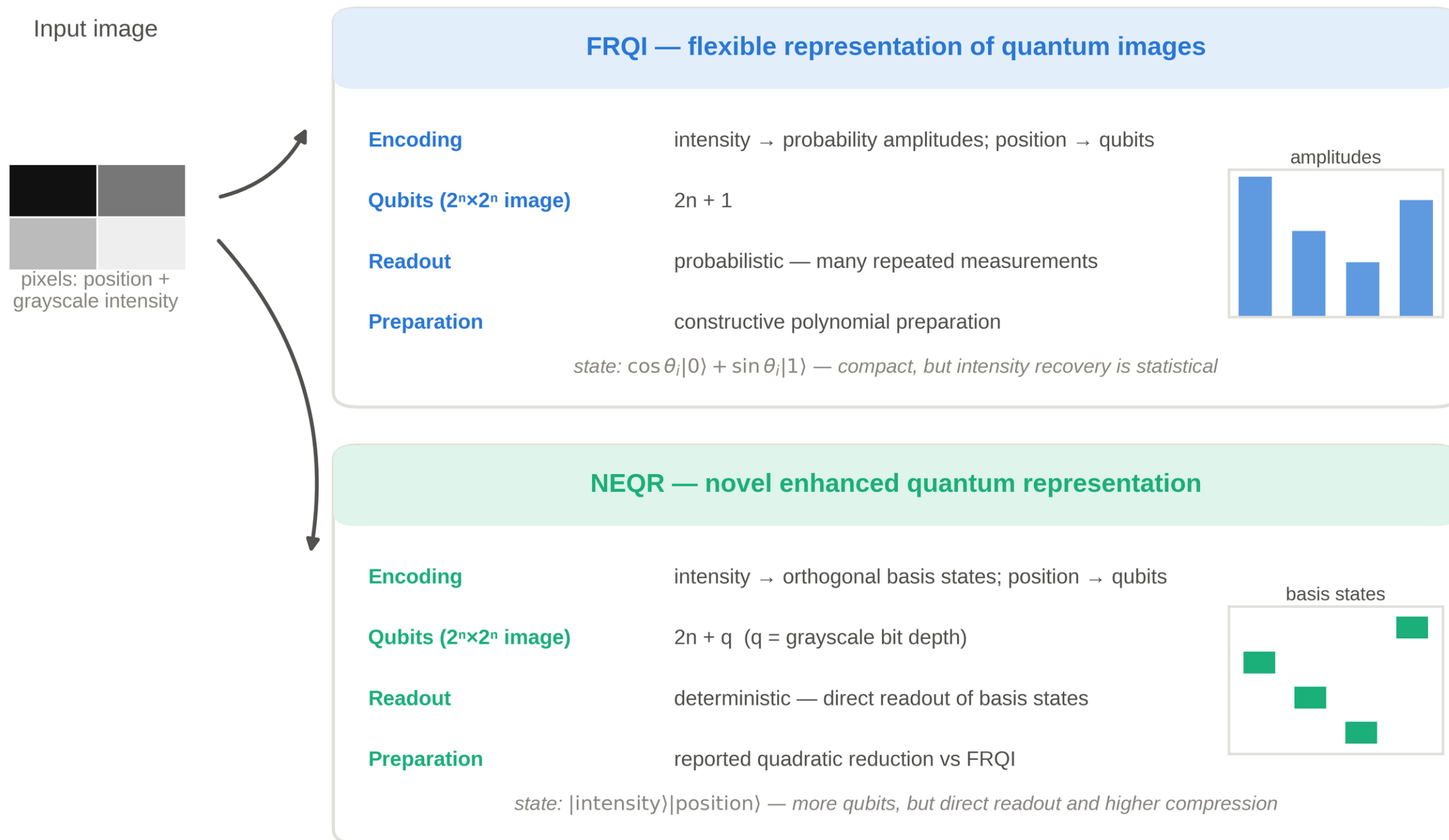

**Figure 5.** Quantum image representations. The flexible representation of quantum images (FRQI) encodes pixel intensities in probability amplitudes, so accurate recovery is probabilistic and requires repeated measurements. The novel enhanced quantum representation (NEQR) stores intensity in orthogonal basis states of a qubit sequence, permitting direct readout and lower reported preparation complexity at the cost of additional qubits.

**Table 4.** Quantum and Quantum-Related Applications by Imaging Modality

| Modality | Main Applications | Representative Findings | Assessment |
|---|---|---|---|
| CT | Reconstruction, segmentation, feature-based classification | Quantum Fourier transform-based algorithms described for reconstruction (8); quantum optimization reconstructed CT images from sinogram patterns in simulation (48) | Potential relevance for low-dose CT and computationally demanding reconstruction |
| MRI | Brain tumor classification and segmentation, volumetric segmentation | Four-qubit QCNN classified 3,264 brain MRI images (89% binary accuracy) (3); hybrid QCNN reached 91.47% validation accuracy (36); reported Dice of 98% on BraTS 2020 requires cautious interpretation (20); quantum annealing selected radiomic features (47) | Broadest evidence base; familiar neuroimaging tasks |
| PET/molecular | Reconstruction, quantum imaging concepts | Quantum Fourier transform-based methods described (8); hybrid adiabatic quantum computing tested on small simulated images (49) | Potential relevance for reconstruction from indirect or incomplete data |
| Ultrasound | Denoising, breast lesion classification | Quantum-inspired denoising reported with signal-to-noise metrics (20); hybrid QCNN on BreastMNIST reported 86.54% test accuracy versus 84.17% for a classical CNN (21) | Clinically intuitive niche; may preserve subtle grayscale features |

Note.—BraTS = Brain Tumor Segmentation challenge, CNN = convolutional neural network, QCNN = quantum convolutional neural network. Numerals refer to the reference list.

### *Image Classification*

Image classification is the most developed QML application. Reported performance represents early proof-of-concept evidence: most studies use benchmark datasets, reduced-resolution inputs, simulator-based experiments, or limited validation splits, so accuracy and area under the receiver operating characteristic curve (AUC) values are best interpreted alongside dataset size and evaluation design (9,24). Most systems are hybrid pipelines: classical feature extraction is usually paired with quantum classification (9,24), and the reverse arrangement, in which quantum convolutional layers feed a classical classifier, has also been reported (35).

QCNNs are a leading model type. A compact four-qubit QCNN classified 3,264 brain MRI images, achieving 89% accuracy for binary tumor detection but 62% for multiclass tumor subtyping, suggesting that current QCNNs are more reliable for simpler binary tasks (3). A hybrid QCNN combining quantum feature-encoding circuits with depthwise-separable convolutions reached 91.47% validation accuracy for brain tumor MRI classification (36). An Inception-v3 transfer learning model feeding a six-layer quantum variational classifier categorized brain MRI as glioma, meningioma, pituitary tumor, or no tumor with per-class accuracy up to 99% on a public dataset, and distinguished high-grade from low-grade glioma with 90.91% accuracy on BraTS 2020 (37).

Chest radiography lends itself to binary quantum classification. A quantum network processing small PneumoniaMNIST patches achieved 83.33% validation accuracy versus 73.33% for a comparable classical CNN (29), and replacing one dense layer of a CNN with a VQC improved AUC (0.836 vs 0.727) across 30 training runs on pediatric radiographs (38). For COVID-19 chest radiographs, a hybrid quantum-classical CNN reported 98.6% binary and 88.2% four-class accuracy (39). Quantum transfer learning, which couples a pretrained classical feature extractor to a small quantum circuit, achieved AUC up to 0.93 for cardiomegaly detection (40), improved stenosis detection accuracy on x-ray coronary angiography by approximately 9% (41), and reached 97.2% accuracy for binary dementia classification on brain MRI (42).

Skin and breast lesions have been studied as additional classification tasks. A quantum support vector classifier built on MobileNet features addressed HAM10000 skin disease classification (43), and a hybrid quantum CNN with a quantum attention mechanism and NEQR encoding reported 91% accuracy for dermoscopic benign-versus-malignant classification (44). A hybrid QCNN has been applied to BreastMNIST (21), and quantum transfer learning reached 84% accuracy on full-field digital mammograms, matching classical transfer learning and exceeding a network trained from scratch (45). Across these settings, QSVMs serve as quantum-enhanced classifiers for compact features rather than as end-to-end image systems, with performance depending on feature selection, encoding strategy, kernel design, and hardware noise (9). Overall, classification is a promising research direction rather than a replacement for deployed radiology AI systems.

### *Image Segmentation*

Current quantum segmentation studies most often use quantum methods as optimization or feature-processing components rather than as fully quantum systems. Brain tumor segmentation is the most developed use case: one reviewed study reported a Dice similarity coefficient of 98% on the BraTS 2020 dataset, although the nonstandard evaluation context warrants cautious interpretation (20). Quantum-inspired methods, algorithms that borrow quantum mathematics but run on classical computers, have improved MRI tumor segmentation by using quantum-based optimization during training, and a qutrit-inspired, fully self-supervised shallow network approached supervised U-Net performance without labeled training data (20,46). A three-dimensional, quantum-inspired, self-supervised model has been evaluated for volumetric segmentation of brain and liver tumors (20). Organ and tissue segmentation remains underdeveloped, with reported examples in brain tissue, hippocampus, and dental radiograph segmentation using quantum-inspired optimization and thresholding (25,32).

### *Radiomics and Feature Extraction*

Medical images contain many quantitative radiomic features, and as feature counts grow, classical models become computationally expensive and prone to overfitting; feature selection reduces high-dimensional data to informative subsets (20). Quantum methods are attractive here because they operate naturally in high-dimensional feature spaces (9). In one brain MRI study, quantum annealing on a D-Wave system selected optimal radiomic features for differentiating large brain metastases from high-grade gliomas, and the resulting explainable quantum neural network matched classical classifiers while using compact, interpretable feature sets (47). Quantum-enhanced modules have also been used to enrich classical imaging features (10). Such approaches may eventually support radiomic characterization of cancer.

### *Image Reconstruction*

Reconstruction converts acquired signal data into interpretable images, often from incomplete or indirect measurements (25); classical deep learning pipelines remain the reference standard, and quantum demonstrations are constrained proofs of concept. Quantum algorithms have been described for MRI, CT, and PET reconstruction, including quantum Fourier transform-based approaches (8,25). Formulating reconstruction as a quantum optimization problem, one study reconstructed CT images from sinogram patterns with high accuracy in simulation (48), and hybrid adiabatic quantum computing has been tested for PET/SPECT reconstruction on small simulated images, with qubit number and connectivity as the limiting factors (49). In single-pixel imaging, quantum neural networks completed reconstruction but with lower quality than classical networks in that experimental setting (50).

### *Applications by Imaging Modality*

Evidence depth varies by modality (Table 4, Figure 6). MRI has the broadest base, particularly in brain tumor classification and segmentation. CT and PET applications center on reconstruction. Ultrasound applications, denoising, and breast lesion classification are limited and clinically intuitive because ultrasound is noisy, low contrast, and operator-dependent (20,21).

## Limitations

Medical imaging applications remain limited by NISQ-era hardware (9,20). Quantum processors require highly controlled operating conditions, making them difficult and expensive to scale for clinical use (11), and noise, decoherence, and high error rates reduce circuit fidelity (9). Deep circuits lose coherence on current devices, mitigation strategies are only partially effective, and gradients of randomly initialized parameterized circuits can vanish exponentially with qubit number (so-called barren plateaus), constraining trainable model depth (7,27).

Small qubit counts restrict task size and force studies to downsample images, compress features, or adopt hybrid or split-circuit designs (7,9,22). Data encoding is a further bottleneck: amplitude encoding is qubit-efficient but circuit-intensive and error-prone, whereas angle encoding is simpler but requires more qubits, limiting achievable image scale (7,9,21).

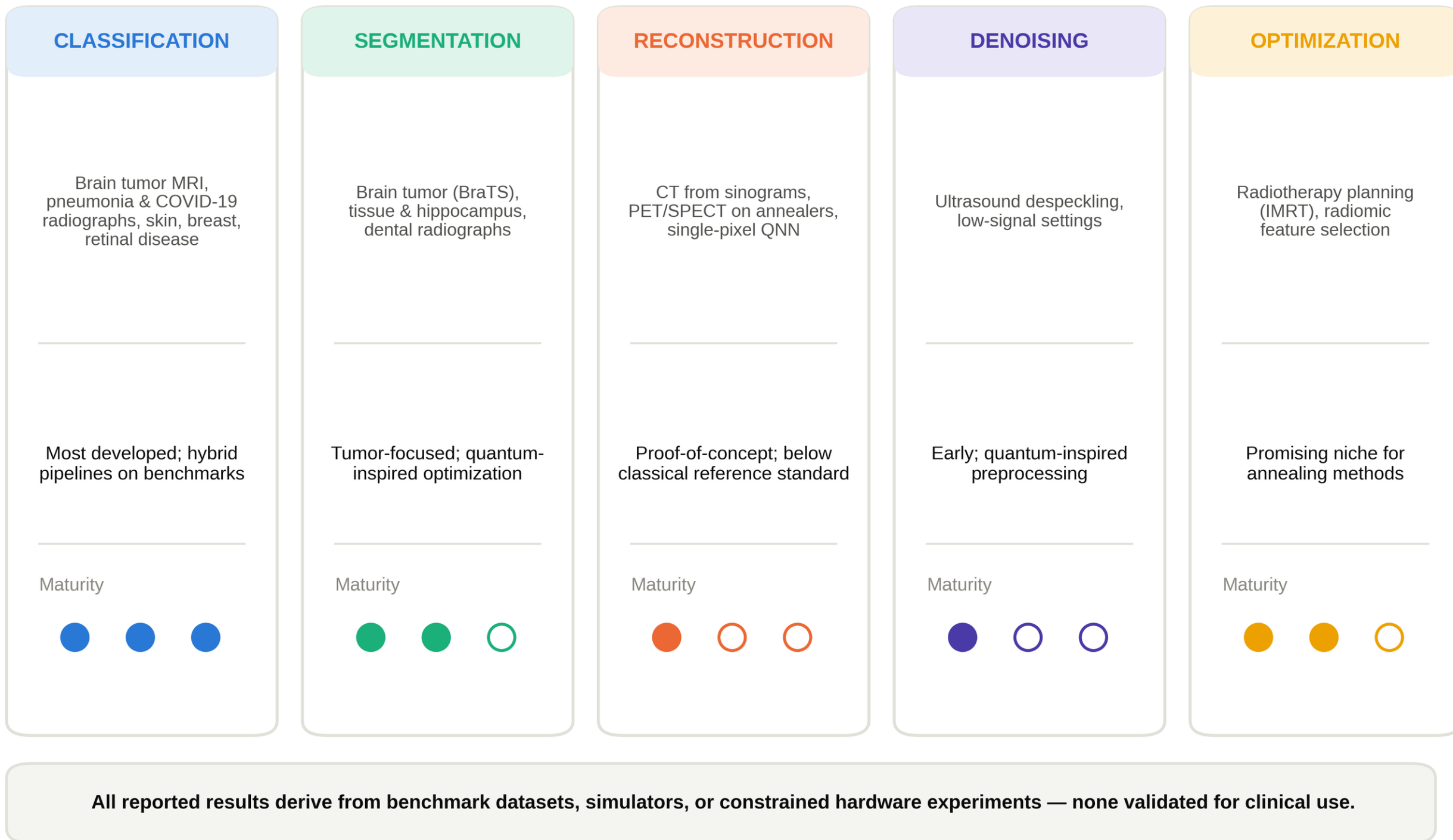

**Figure 6.** Application landscape of quantum and quantum-inspired methods in medical imaging: classification (most developed), segmentation (tumor-focused), reconstruction (early CT and PET proofs of concept), denoising (ultrasound and low-signal settings), and optimization (radiotherapy planning and radiomic feature selection). All reported results derive from benchmark datasets, simulators, or constrained hardware experiments.

## Conclusion

Quantum computing in radiology remains in its early stages and is predominantly experimental. Most studies are hybrid or proof-of-concept systems limited by current hardware rather than routine clinical tools. The early signal is promising, particularly for hybrid models in selected classification tasks, but clinical value is unproven until studies move beyond simulators, small datasets, and weak or inconsistent baselines. Translating promise into clinically relevant impact will require close collaboration among radiologists, physicists, and computer scientists, so that advances in quantum algorithms and hardware remain anchored in diagnostic problems and radiology practice (19).

## Declarations

**Funding:** This research did not receive any specific grant from funding agencies in the public, commercial, or not-for-profit sectors.

**Conflicts of interest:** The authors declare that they have no conflicts of interest relevant to this work.

**Artificial intelligence disclosure:** During the preparation of this work, the authors used Grammarly AI to assist with improving grammar and syntax. The authors critically reviewed and edited all outputs and take full responsibility for the publication's content.